\documentstyle[prl, twocolumn, aps, psfig]{revtex}

\begin{document}
\draft
\preprint{}
\title{Irrotational Binary Neutron Stars in Quasi-Equilibrium}

\author{P. Marronetti and G. J. Mathews}

\address{
University of Notre Dame,
Department of Physics,
Notre Dame, Indiana 46556}

\author{J. R. Wilson}

\address{
University of California,
Lawrence Livermore National Laboratory,
Livermore, California  94550}

\date{\today}
\maketitle
\begin{abstract}
We report on numerical results from an independent formalism to describe
the quasi-equilibrium structure of nonsynchronous binary neutron stars in general 
relativity. This is an important independent test of controversial numerical 
hydrodynamic simulations which suggested that nonsynchronous neutron stars in a close
binary can experience compression prior to the last stable circular orbit. We show 
that, for compact enough stars the interior density increases slightly as irrotational
binary neutron stars approach their last orbits. The magnitude of the effect, however, is 
much smaller than that reported in previous hydrodynamic simulations.
\end{abstract}

\narrowtext

\section{Introduction}

The physical processes occurring during the last orbits of a neutron-star 
binary are a subject of intense current debate 
\cite{wm95,wmm96,mw97,mmw98,lai,others,flanagan,thorne97,baumgarte,sbs98,Bon99}.
In part, this recent surge in interest stems from relativistic numerical 
hydrodynamic simulations in which it has been noted \cite{wm95,wmm96,mw97} 
that as the stars approach each other their interior density increases.
Indeed, for an appropriate mass and equation of state, previous 
numerical simulations indicated that binary neutron stars might collapse 
individually toward black holes seconds prior to merger. This compression 
effect would have a significant impact on the anticipated gravity-wave signal 
from merging neutron stars and may also provide an energy source for cosmological 
gamma-ray bursts \cite{mw97}.
 
In view of the  unexpected nature
of this neutron star compression effect, and its possible repercussions,
as well as the extreme complexity of strong field general relativistic hydrodynamics,
it is of course imperative that there be an independent confirmation of the
existence of neutron star compression before one can  be convinced
of its operation in binary systems.  This is particularly important
since many authors \cite{lai,others,flanagan,thorne97,baumgarte,sbs98}
have searched for but not observed this effect in Newtonian tidal forces 
\cite{lai}, first post-Newtonian (1PN) dynamics \cite{others,sbs98},
tidal expansions \cite{flanagan,thorne97}, or in binaries
in which rigid corotation has been imposed \cite{baumgarte,pmw98}.
In ref. \cite{mmw98} it has been argued that none of the above works could or 
should have observed the effect, since the compression only dominates over 
tidal forces for stars with a realistic compaction ratio [$(M_G/R)_\infty$]
that are in a nearly irrotational hydrodynamic state 
(i.e. little spin relative to a distant observer). Indeed, irrotational 
stars may be a likely configuration near the final orbits as corotation 
would demand an unrealistically large viscosity in neutron stars \cite{bc92}. 
In hydrodynamic relaxation calculations \cite{mmw98} the stars even seem 
to prefer a nearly irrotational state. Rasio and Shapiro \cite{Ras99} provide a
recent review of the state of the aforementioned controversy.

With this in mind it is of particular interest that a new formalism  
\cite{bonazzola,teukolsky,shibatab,asada,gourgoul}
has been proposed in which the hydrostatic quasi-equilibrium of irrotational 
stars in a binary can be solved independently of the complexities of (3+1) 
numerical relativistic hydrodynamics. Thus, this provides an opportunity to
independently test the hydrodynamics result. Previous results using this formalism have
been presented by Bonazzola {\it et al.} \cite{Bon99}. They show that for an irrotational
binary system composed of neutron stars with $(M_G/R)_\infty$ of 0.14 there is
almost no evidence of the compression effect (although they did note that
the central density for irrotational stars remains much higher than for corotating
stars as the orbit decays). In this paper we report results for two
different systems involving stars with $(M_G/R)_\infty$ of 0.14 and 0.19.
We show that, while the former sequence shows almost no compression in agreement 
with \cite{Bon99}, stars with a higher compaction ratio seem to experience a slight 
central compression as the stars approach.

We also note that these sequences have an important intrinsic value beyond the 
controversy about the
compression effect. They provide realistic solutions to the initial value problem for
neutron-star binary systems that can be used as starting points of fully dynamical 
relativistic hydrodynamical simulations. The fact that they are valid in the strong
field regime makes them more attractive than post-Newtonian counterparts, since
the simulations can be initiated at stages very close to the final merger of the stars.
	
\section{The method}
 
The method we use to determine the internal structure of stars in
irrotational quasi-equilibrium configurations is essentially that 
originally proposed by Bonazzola, Gourgoulhon \& Marck \cite{bonazzola} 
and as simplified by Teukolsky \cite{teukolsky}. The derivation 
of the relevant equations can be found in those papers. The essential 
ingredient of this approach is that, if the fluid vorticity is zero 
(e.g. as in irrotational stars), the specific momentum density per baryon 
can be written as the gradient of a scalar potential,
\begin{equation}
h u_\mu = \nabla_\mu \psi~~,
\end{equation}
where $u_\mu$ is the covariant four velocity and $h$ is the relativistic enthalpy,
$h = 1 + \epsilon + {P \over \rho_0}$,
where $\epsilon$ is the internal energy per unit of baryon rest mass, 
$P$ is the pressure, and $\rho_0 = m_B n_B$  is the proper baryon rest mass density, 
with $n_B$ the baryon number density. The potential $\psi$ can be obtained 
from the solution of a Poisson-like equation:
\begin{equation}
D^i D_i \psi = D_i {\lambda B^i \over \alpha^2}
- \biggl(D^i \psi - {\lambda \over \alpha^2} B^i \biggr)
D_i~ln{\biggl({\alpha n_B \over h}\biggr)}~~,
\label{psi}
\end{equation}
where  $D_i$ are spatial covariant derivatives, and
\begin{equation}
\lambda =  C + B^j D_j \psi  = \alpha[h^2 + (D_i \psi)^2]^{1/2}~~,
\end{equation}
where $C$ is a constant.  The quantities $\alpha$ and $B^i$ are obtained 
from the usual ADM (3+1) metric
\begin{equation}
ds^2 = -(\alpha^2 - \beta_i\beta^i) dt^2 + 2 \beta_i dx^i dt + \gamma_{ij}dx^i dx^j~~,
\end{equation}
such the $B^i$ is the shift vector in the rotating frame,
$B^i = \beta^i + (\omega \times r)^i$,
where $\omega$ is the angular velocity of orbital motion.

Equation (\ref{psi}) must be solved by imposing a boundary condition
at the stellar surface,
\begin{equation} 
\biggl( D^i \psi - {\lambda \over \alpha^2} B^i \biggr) D_i n_B\Big \vert_{surf} = 0~~.
\label{psi_bc}
\end{equation}
For irrotational stars the Bernoulli integral for the matter distribution
then becomes:
\begin{equation}
h^2 = -(D^i \psi) D_i \psi + {\lambda^2  \over \alpha^2}~~.
\label{bernoullib}
\end{equation}
Equation (\ref{bernoullib}) uniquely determines the equilibrium structure of 
the stars.

We also compute stars in constant corotation. In this case the relativistic 
Bernoulli equation can be written \cite{baumgarte,pmw98} 
${h / u^0}  =  {\rm constant}$.

In the present work we consider a polytropic equation of state, 
$ P = K \rho_0^\Gamma$, with $\Gamma = 2$ and 
$K= 1. 13\times 10^5$ erg cm$^3$ g$^{-2}$. This gives a maximum neutron-star 
gravitational  mass of 1.46 $M_\odot$. In these simulations we consider two
equal-mass neutron stars in two different sequences of stable orbits.
The first corresponds to stars with baryon mass of $M_B = 1.29~M_\odot$,
gravitational mass in isolation of $M_{G\infty} = 1.21~M_\odot$, and compaction ratio
of $(M_G/R)_\infty = 0.14$. This sequence is comparable to the one studied by
Bonazzola {\it et al.} \cite{Bon99}. The second sequence corresponds to a systems
of more compact stars, with  baryon mass of $M_B = 1.55~M_\odot$,
gravitational mass in isolation of $M_{G\infty} = 1.41~M_\odot$, and compaction ratio
of 0.19. For the grid resolution of this study ($\sim 40$ zones 
across the star) we obtain a proper central density in isolation of
$\rho_\infty =0.96\times 10^{14}$ g cm$^{-3}$ for the first sequence and 
$\rho_\infty =1.83\times 10^{14}$ g cm$^{-3}$ for the second one. 
As pointed out in \cite{mmw98} it is important to study realistically compact neutron 
stars. Otherwise Newtonian tidal forces can dominate over the relativistic effects one 
desires to probe.  

The Einstein field equations are solved by imposing a conformally flat condition 
(CFC)  on the three metric \cite{wm95,wmm96}. That is, the spatial three 
metric is constrained to be represented by a position dependent conformal 
factor $\phi^4$ times the Kronecker delta, $\gamma_{i j} = \phi^4 \delta_{i j}$.
This is a common metric  choice for solving the initial value problem in 
numerical relativity.  It is consistent with the approximation of quasi-equilibrium 
circular orbits \cite{mmw98,teukolsky} which we wish to evaluate
(see ref. \cite{mmw98} for a detailed discussion).

The advantage of this method is that the determination of the metric
coefficients reduces to the solution of flat-space Poisson like
equations  \cite{wmm96}. For example, using the Hamiltonian constraint
\cite{y79} in combination with the maximal slicing condition 
$tr(K)=0$, the equation for $\phi$ becomes
\begin{equation}
\nabla^2{\phi} = -4\pi{\phi^5 \over 2}\biggl[ \rho_0 hW^2 -  P
+ {1 \over 16\pi} K_{ij}K^{ij}\biggr]
~~,
\label{phi}
\end{equation}
where, $W \equiv  \alpha u^0$.  A similar equation can be written \cite{wmm96}
for the lapse function.

The shift vector $\beta^i$ is further  decomposed \cite{Bow80}:
$\beta^i = G^i - (1 / 4)\nabla^i \chi$, and introduced into the ADM 
momentum constraint equation to obtain two elliptic equations
\begin{equation}
\nabla^2 \chi    = \nabla_i G^i~~~,
\label{chibeta}
\end{equation}
\begin{equation}
\nabla^2 G^i    = 2 \nabla_j ln(\alpha \phi^{-6}) K^{ij} 
- 16 \pi \alpha \phi^4 S^i ~.
\label{capb}
\end{equation}
These equations have been solved with boundary conditions provided by
the first terms in a multipole expansion of the fields as described in 
\cite{pmw98}. 
An equation for the extrinsic curvature $K^{ij}$ follows \cite{wmm96} from 
the ADM momentum constraint equation with the maximal slicing condition.

This set of equations (\ref{psi}), (\ref{bernoullib})-(\ref{capb})
is solved numerically using an iterative algorithm based upon a specially 
designed elliptic solver. This method consists of a combination of multigrid 
algorithms and domain decomposition techniques \cite{pmw98,Mar99} and utilizes 
a code which was developed independently of the hydrodynamics code of \cite{wm95,wmm96}.

Solutions are obtained for specific values of the coordinate distance between 
stars and the total baryonic mass. In this way we can construct a constant
baryonic-mass sequence of orbits with a minimum number of code runs. This 
sequence is a collection of semi-stable orbits that are connected by the 
inspiral motion of the stars.


The problem consists essentially in the numerical solution of a set of elliptic 
equations. The cases of corotating and irrotational systems share the same set 
of equations for the metric fields.
The irrotational systems, however, demand the solution of an extra elliptic 
equation (\ref{psi}) for the description of the stellar internal structure.
This poses a very special problem since the boundary condition
(\ref{psi_bc}) is to be satisfied on the stellar surface and not 
on the grid boundaries as for the rest of the elliptic equations. This is 
challenging to implement numerically on a fixed Eulerian grid, since the stellar 
surface is 
a spheroid embedded in a Cartesian grid. We refer the reader to \cite{Mar99}
for the details of our approach to this calculation. Note, however, that this is a completely 
different numerical approach to that of Bonazzola {\it et al.} who utilized spectral
methods to solve the elliptic equations. 

\begin{figure}[htb]
\begin{center}
\hskip 2.0 cm
\vskip .5 cm
\mbox{\psfig{figure=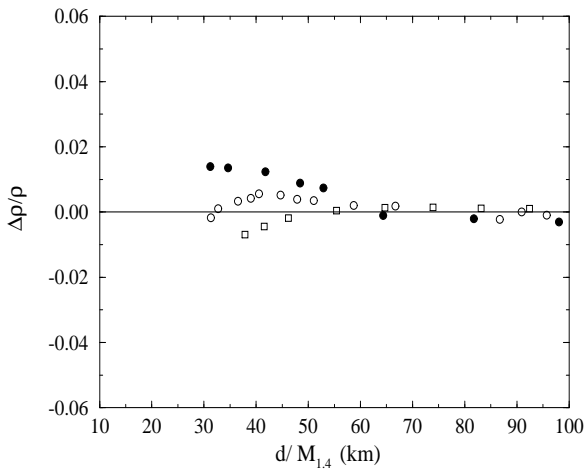,width=6 cm,height=5 cm}}
\end{center}
\vskip 0.5 cm
\caption{Change in central density relative to a single isolated star
$(\Delta \rho/\rho)$ as a function of the coordinate distance. 
The black (white) circles represent our calculations for 
sequences with $(M_G/R)_\infty$ = 0.19 (0.14). The squares
show the result obtained by Bonazzola {\it et al.} for a sequence composed of stars 
with $(M_G/R)_\infty$ = 0.14.}
\label{drho}
\end{figure}

To quantify the numerical accuracy, the code was tested against the Newtonian 
irrotational sequences obtained by
Ury\={u} and Eriguchi \cite{uryu} in two different
ways. In one test, the code was stripped of all relativistic 
terms to reduce it to Newtonian physics. 
In a second test, orbits for low mass stars ($M_B = 0.10$ $M_\odot$) were
calculated using the relativistic code to approach 
the Newtonian regime ($M_G/R \approx 0.01$).
Comparisons were done for orbits with fixed separation distances between stellar
centers ($\tilde{d}$ in the units of \cite{uryu}).  
Most importantly, both of these tests
exhibit the Newtonian expectation of no discernible change in 
central density relative to that of isolated stars, with ($\Delta \rho/\rho
~^<_\sim  0.1\%$).
The resulting values (and percentage difference from results of \cite{uryu}) 
 for total energy, total angular momentum, 
and orbital angular frequency were  as follows: For the first test, 
$\tilde{d} = 3.53$, and 
 $\tilde{E} = -1.142~(0.1\%)$, $\tilde{J} = 1.278~(4\%)$, and 
$\tilde{\Omega} = 0.241~(3\%)$;  In the 
second test, $\tilde{d} = 3.88$, and
$\tilde{E} = -1.082~(4\%)$, $\tilde{J} = 1.413~(1\%)$, and $\tilde{\Omega} =
0.203~(6\%)$. These results are consistent with the expected numerical
accuracy of this comparison. The code was also tested simulating corotating binary
orbits (see \cite{pmw98}).

Finally, Flanagan \cite{Fla99} pointed out an inconsistency in the definition of 
momentum density in the original hydrodynamics calculations \cite{wm95,wmm96,mw97}.
This problem is not present in these calculations.

\section{\bf Results}

For the present study we have found solutions to the initial value 
equations described above for semi-stable circular orbits
for a binary system of identical neutron stars extending from
the post-Newtonian regime to the innermost orbit for which
we obtain a stable solution.

Figure \ref{drho} shows the fractional change $\Delta \rho/\rho$ in the 
proper central rest-mass density relative to the central density $\rho_\infty$ 
of an isolated star of the same resolution. Results are plotted as a 
function of the normalized coordinate distance; i. e. 
$d/M_{1.4} \equiv d/ (M_G/{1.4~M_\odot})$.

\begin{figure}[htb]
\begin{center}
\hskip 2.0 cm
\vskip .5 cm
\mbox{\psfig{figure=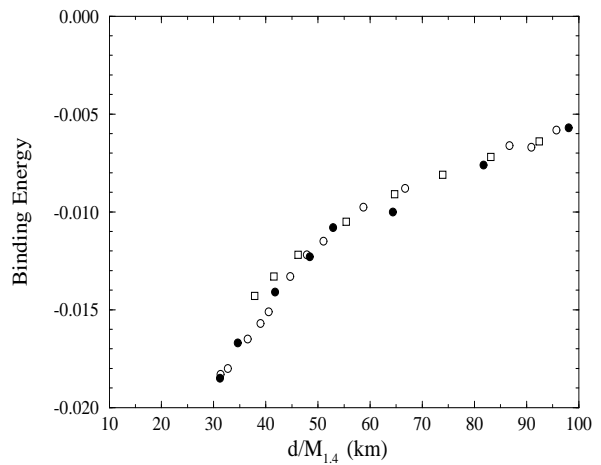,width=6 cm,height=5 cm}}
\end{center}
\vskip 0.5 cm
\caption{Binding energy defined as $(M_G-M_{G\infty})
/M_{G\infty}$, where $M_G$ is half the ADM mass of the system, vs coordinate distance.
The black (white) circles correspond to the sequence of stable orbits with 
$(M_G/R)_\infty$ = 0.19 (0.14). The squares
show the result obtained by Bonazzola {\it et al.} for a sequence composed of stars 
with $(M_G/R)_\infty$ = 0.14.}
\label{BE}
\end{figure}

The squares show the orbits obtained by Bonazzola {\it et al.}
\cite{Bon99} for a system composed of stars with $(M_G/R)_\infty$ = 0.14. 
The black (white) circles represent our calculations for sequences with 
$(M_G/R)_\infty$ = 0.19 (0.14). The numerical error 
in $\Delta \rho/\rho$ for our irrotational points is 
conservatively estimated to be $\sim  \pm 0.005$ based upon the 
code tests here described as well as various convergence tests.

Note, that both the Bonazzola {\it et al.} results and ours for $(M_G/R)_\infty$ = 0.14
show only a very slight increase in the central density before the tidal effects begin 
to dominate at short separations. The difference between the two curves is probably
attributable to differences in the numerical algorithms applied
(i.e.~ coordinate mesh, boundary conditions, elliptic solver, etc). 
Nevertheless they are consistent to within numerical accuracy.
We emphasize that our point is not about quantitative details at the
level of $\pm 0.005$ in $\Delta \rho/\rho$, but about the 
qualitative trend of increasing central density as the compaction ratio increases.

The sequence with $(M_G/R)_\infty$ = 0.19,
clearly, raises above the zero line, ending in a maximum value of $\sim 1.5\%$ 
increase in the central density for the last stable orbit found. 
This result agrees with a similar trend of increasing compression
with increasing compaction ratio, presented recently by Bonazzola {\it et al.} \cite{Bon98}.
This trend derives from the fact that the compression effect competes with the 
tidal deformation and the latter is stronger for more extended (i.e. smaller 
compaction ratio) stars.
 
In both cases the central density approaches the isolated-star limit at 
large orbital separations. We have checked that the same trend emerges 
in a plot of the average density, so this effect could not be an artifact 
of the stellar center being a special point.

As a final point, figure \ref{BE} shows the relative binding energy of the system 
defined as $(M_G-M_{G\infty})/M_{G\infty}$, where $M_G$ is half the ADM mass of the 
system. We note that the 
relative binding energy does not depend strongly on the compaction ratio
within the numerical accuracy of our results. 
Another feature of these curves is that they lack a turning point in the sequence. 
This is in qualitative agreement with the Newtonian irrotational sequences for 
polytropic index $n=1$ obtained by Ury\={u} and Eriguchi \cite{uryu}.

In summary, the present independent study has obtained the qualitative 
result of increasing central density as an irrotational binary orbit
decays, for neutron stars with compaction ratio of 0.19. The magnitude of the effect,
however, is significantly less than that of the previously reported hydrodynamic
results \cite{wm95,wmm96,mw97}. Preliminary recent results \cite{mwb} indicate that
such reduction in compression effect is consistent with the effect of a correction
proposed by Flanagan \cite{Fla99} when applied to the hydrodynamic simulations.

Although it is evident that some compression effect exists, it is not yet
clear whether this remaining effect is real or a consequence of the 
conformally flat metric approximation.  The magnitude of the apparent
effect noted here
is roughly comparable to the possible uncertainty introduced by this approximation
(cf.~\cite{mmw98}).  The ultimate test of whether the effect is real will
therefore require an accurate fully dynamical relativistic treatment.
Such a test is hopefully coming from the neutron-star Grand Challenge
collaboration.

 
Work at University of Notre Dame supported by NSF grant PHY-97-22086. Work at the 
Lawrence Livermore National Laboratory performed in part under the auspices of the 
U.~S.~Department of Energy under contract W-7405-ENG-48 and NSF grant PHY-9401636.

\end{document}